\documentclass[a4paper]{article}

\usepackage{multirow}
\usepackage{array}

\usepackage{INTERSPEECH2019}
\usepackage{adjustbox}
\usepackage{hyperref}

\title{STC Speaker Recognition Systems for the VOiCES From a Distance Challenge}
\name{Sergey Novoselov$^{1, 2}$, Aleksei Gusev$^{1, 2}$, Artem~Ivanov$^1$, Timur Pekhovsky$^1$, Andrey Shulipa$^2$,  \\ Galina Lavrentyeva$^{1, 2}$,  Vladimir Volokhov$^1$, Alexandr~Kozlov$^1$}
\address{
  $^1$STC-Innovations Ltd., St. Petersburg, Russia \\
  $^2$ITMO University, St. Petersburg, Russia}
\email{\{novoselov, gusev-a, ivanov-ar,  tim, shulipa,  \\  lavrentyeva, volokhov, kozlov-a\}@speechpro.com}

\begin{document}

\maketitle
\begin{abstract}
This paper presents the Speech Technology Center (STC) speaker recognition (SR) systems submitted to the VOiCES From a Distance challenge 2019 \footnote{This work was partially financially supported by the Government of the Russian Federation (Grant 08-08).}. The challenge's SR task is focused on the problem of speaker recognition in single channel distant/far-field audio under noisy conditions. In this work we investigate different deep neural networks architectures for speaker embedding extraction to solve the task. We show that deep networks with residual frame level connections outperform more shallow architectures. Simple energy based speech activity detector (SAD) and automatic speech recognition (ASR) based SAD are investigated in this work. We also address the problem of data preparation for robust embedding extractors training. The reverberation for the data augmentation was performed using automatic room impulse response generator. In our systems we used discriminatively trained cosine similarity metric learning model as embedding backend. Scores normalization procedure was applied for each individual subsystem we used.
Our final submitted systems were based on the fusion of different subsystems.
The results obtained on the VOiCES development and evaluation sets demonstrate effectiveness and robustness of the proposed systems when dealing with distant/far-field audio under noisy conditions.


\end{abstract}

\section{Introduction} \label{sec:intro}

Text-independent speaker recognition remains a challenging task for modern voice biometrics systems. Complex speaker voice information must be captured from highly variable data with no evident speaker patterns. Candidate solutions must generalize well in order to be robust to new possible deployment conditions. 

The last investigations performed for NIST SRE 2016 \cite{sre_2016} and NIST SRE 2018 \cite{NIST2018SREP} datasets confirm that discriminatively trained deep speaker embeddings extractors provide State-of-the-Art performance in SR task.
According to the results of previous studies on text-independent speaker recognition in telephone \cite{sre_2016} and microphone channels \cite{SITW_ep}, deep speaker embeddings based systems (like x-vectors) significantly outperform conventional i-vector based systems in terms of speaker recognition performance.
In addition, recent studies \cite{NSKKS2018, huang2018angular,cai2018exploring} present the successful implementation of some proven approaches from face recognition field for deep speaker embeddings extractors training.
A comparative study of different back-end solutions for DNN based speaker embeddings was presented in \cite{NSSKK2018}. This work demonstrated that cosine similarity metric learning (CSML) approach can be effectively used for speaker verification in deep neural network (DNN) embeddings domain. It was shown that the performance of deep speaker embeddings based systems can be improved by using CSML with the triplet loss training scheme in both ”clean” and ”in-the-wild” conditions.

The VOiCES from a Distance challenge 2019 (VOiCES~2019 challenge) \cite{richey2018voices, nandwana2019voices} is aimed to support research in the area of speaker recognition and automatic speech recognition with the special focus on single channel far-field audio under noisy conditions. 

This paper describes the STC speaker recognition systems submitted to the VOiCES challenge for both fixed and open conditions.
During the challenge, we explored several systems based on deep speaker embeddings extraction. With the example of some systems from NIST 2018, we considered deeper neural networks architecture with additional recurrent layers (LSTM) on the frame level.

In this work we also address the problem of data preparation for robust embedding extractors training. In particular, we used automatic room impulse response (RIR) generator to simulate reverberation during the data augmentation process.
Additionally, we explored different speech activity detectors. Simple energy based speech activity detector (SAD) and automatic speech recognition (ASR) based SAD were investigated.

Taking into account the results of \cite{NSSKK2018} we analyse the efficiency of CSML as a Back-End scoring model.
The rest of the paper contains descriptions and implementation details of the systems submitted to  the VOiCES 2019 challenge.




\section{System components description} \label{sec:systems}
In this section we provide a description of all the components used in our systems.

\subsection{Front-End}

Two types of Mel Frequency Cepstral Coefficients (MFCC) were used in this research as low-level features:

\begin{itemize}
\item 23 dimensional MFCC extracted from raw audio signal (8000 Hz) with 25ms frame-length and 15 ms overlap;
\item 40 dimensional MFCC extracted from raw audio signal (16000 Hz) with 25ms frame-length and 15 ms overlap.
\end{itemize}

After the features were extracted we applied two different postprocessing techniques depending on the type of embedding extractor used later:
\begin{itemize}
    \item local Cepstral Mean Normalization (CMN) over a 3-second sliding window;
    \item local CMN over a 3-second sliding window and global Cepstral Mean and Variance Normalization (CMVN) over the whole utterance.
\end{itemize}

We explored two types of speech activity detectors for SR task: energy-based SAD from Kaldi Toolkit and a more sophisticated ASR-based SAD \cite{MedennikovVOiCES2019}. 

Weighted Prediction Error (WPE) method was used in some cases before MFCC extraction for speech dereverberation in order to improve signal quality. We used the open-source implementation\footnote{\url{https://github.com/fgnt/nara\_wpe}}~\cite{Drude2018NaraWPE} of the WPE~\cite{NYKMJ2010} algorithm.

\subsection{Speaker embedding}
In this work we focused on two types of deep neural network speaker embeddings: x-vectors \cite{SGRSPK2018} and Speaker Residual Net based embeddings recently proposed by the authors \cite{NSKKS2018}. We refer to the latter as c-vectors.

All STC x-vector systems for this challenge utilized Kaldi Toolkit \cite{povey2011kaldi}.
Our x-vector systems were mainly based on the configuration described in \cite{SGRPK2017} and its modifications (see Section \ref{xfixed}). The speaker embeddings in this case are extracted from the affine layer on top of the statistics pooling layer of the classifier network.


All STC c-vector systems for this challenge utilized Pytorch \cite{paszke2017automatic}.
Our c-vector systems were mainly based on the configuration described in \cite{NSKKS2018, NSSKK2018}. In addition to the ideas outlined in these papers we used factored form of TDNN for some of c-vector extractors implementation.

To train speaker embedding extractors we used different dataset configurations described in Section~\ref{sec:data_part}. Moreover, we explored different MFCC configurations and SAD systems in our subsystems. A detailed description of the extractors is presented in Section~\ref{sec:implementation}.

\subsection{Back-End}

To discriminate speakers in a DNN-based  speaker embeddings space we used CSML approach with the triplet loss training scheme~\cite{NB2010}.
Scores s-normalization technique from \cite{CVDFKCL2017} was used for both x-vector and c-vector systems.


\subsection{Fusion and Calibration}
\label{fusion_calibration}
In our experiments the development set was divided into two equal parts (devset-I and devset-II). One was used for scores normalization and the other was used for calibration and fusion parameters tuning and vice versa.

The final STC submission systems were obtained by the fusion of several subsystems at the score level.  Fusion was performed by a linear regression model with equal weights for all the individual systems or weights obtained empirically (based on the individual systems quality estimation).

The calibration of the fused systems was done by logistic regression using the BOSARIS toolkit \cite{BV2013}.


\section{Training data} \label{sec:data_part}
According to the last studies \cite{SGRPK2017, mclaren2018train, NSKKS2018} training data preparation plays a crucial role in deep speaker embeddings extractor training. Therefore, in this work we paid great attention to the selection of training data for tuning speaker recognition systems in single channel far-field audio, under noisy conditions. 
\subsection{Fixed training condition}
The recordings sampling rate in fixed conditions was 16000 Hz. We considered the following three versions of the training set for fixed conditions:

\textbf{FixData-I} includes VoxCeleb1, VoxCeleb2 (development set) and their augmented versions. Augmented data was generated using standard augmentation recipe from Kaldi Toolkit \cite{SGRSPK2018} (reverberation, babble, music and noise) using the freely available MUSAN and RIR datasets \footnote{ \url{http://www.openslr.org}}.  Augmentation was performed in order to simulate the distortions typical to far-field microphone under noisy conditions. Reverberation was added to both clean and distorted (babble, music and noise) sound recordings. The final database consists of approximately 5,600,000 examples (7205 speakers). Energy-based SAD from Kaldi Toolkit \cite{SGRSPK2018} was applied to select speech frames from the data. Audio samples with speech duration less than 3.5 seconds were excluded and the maximum amount of samples for one speaker was limited to 8. 

\textbf{FixData-II}\footnote{ We would like to thank the STC\_ASR team that participated in speech recognition task of the VOiCES Challenge \cite{MedennikovVOiCES2019} for their help with data augmentation, ASR SAD implementation and helpful discussion.} consists of VoxCeleb1, VoxCeleb2 and SITW and their augmented versions. The augmented data were obtained in a way similar to FixData-I, but reverberation was performed using the impulse response generator based on \cite{AB1979}. Four different RIRs were generated for each of 40,000 rooms with a varying position of sources and destructors. It should be noted that, in contrast to the original Kaldi augmentation, we reverberated both speech and noise signals. In this case different RIRs generated for one room were used for speech and noise signals respectively. Thus we obtained more realistic data augmentation. The final database consists of approximately 5,200,000 examples (7562 speakers). Similarly to FixData-I, energy-based SAD  \cite{SGRSPK2018} was applied to filter out nonspeech frames.

\textbf{FixData-III}: This database is similar to FixData-II, but ASR based SAD \cite{MedennikovVOiCES2019} was used to preprocess the examples from the database instead of the energy-based SAD.

\subsection{Open training condition}
We extended the training dataset for open conditions by adding telephone channel data from NIST SREs datasets.

\textbf{OpenData-IV}: All data from the  NIST 2018 SRE fixed training conditions with VoxCeleb1, VoxCeleb2 (development set) and SITW.  Augmented data was generated using  standard augmentation from Kaldi Toolkit \cite{SGRSPK2018} (reverberation, babble, music and noise). Energy-based SAD from Kaldi Toolkit \cite{SGRSPK2018} was applied to preprocess the examples from the database. The final database consists
of approximately 8,000,000 examples (13613 speakers). All data was downsampled to 8000 Hz.


\newcolumntype{K}[1]{>{\raggedright\arraybackslash}m{#1}}
\newcolumntype{C}[1]{>{\centering\arraybackslash}m{#1}}

\section{Implementation details} \label{sec:implementation}

\subsection{Fixed training conditions}
\label{common_fixed}
\subsubsection{X-vector based systems}
\label{xfixed}

All considered x-vector based systems for fixed conditions utilize 40 dimensional MFCC with local CMN-normalization as input features and CSML as a back-end.

\textbf{Xvec-TDNN-V1}: Standard x-vector system described in \cite{SGRSPK2018}. 
FixData-I was used to train the embedding extractor.

\textbf{Xvec-TDNN-V2}: Xvec-TDNN trained on FixData-II.

\textbf{Xvec-TDNN-V3}: Xvec-TDNN trained on FixData-III.

\textbf{Xvec-TDNN-LSTM-V1}: This system configuration is based on Xvec-TDNN. The difference is that the 4th layer was replaced by LSTM layer with cell dimension of 512, delay in the recurrent connections equal to -3, and both recurrent and non-recurrent projection dimension equal to 256. The embedding extractor was trained on FixData-I.

\textbf{Xvec-TDNN-LSTM-V3}: Xvec-TDNN-LSTM trained on FixData-III.

\textbf{Xvec-Ext-TDNN-V1}: Configuration of this system was an extended version of the original TDNN system used in Xvec-TDNN-V1. The differences here include an additional TDNN layer with wider temporal context, and unit context TDNN layers between wide context TDNN layers. This approach is taken from the JHU-MIT System Description for NIST SRE18 \cite{NIST2018SREP}. This embedding extractor was trained on FixData-~I.

\textbf{Xvec-Ext-V3}: Xvec-Ext-TDNN trained on FixData-III.

\textbf{Xvect-Ext-TDNN-LSTM-V3}: This system is the extended version of the original x-vector extractor, similar to Xvec-Extended-TDNN, but with 9th layer replaced by LSTM-layer with cell dimension of 512, delay in the recurrent connections equal to -3, and both recurrent and non-recurrent projection dimension equal to 256. The LSTM layer context was reduced to 3. This embedding extractor was trained on FixData-III.

\textbf{Xvect-FTDNN-V2}: This system is based on the factorized TDNN embedding extractor system proposed by JHU-MIT for NIST SRE18 \cite{NIST2018SREP}. The main idea is that TDNN pre-pooling layers in the original X-vector system are replaced by factorized TDNN with skip connections, where prior layers are concatenated to the input of the current layer. The proposed system mainly differs from the original FDNN one in changed skip connections and reduced sizes of TDNN layers. More architecture details are presented in Table~\ref{tab:table_ftdnn}. This configuration contained less parameters than Extended TDNN based systems. FixData-II was used for training.

\begin{table}[h]
\centering
\caption{Factorized TDNN configuration}
\label{tab:table_ftdnn}
\resizebox{\columnwidth}{!}{
\begin{tabular}{|l|K{2cm}|K{1cm}|K{1cm}|C{1cm}|C{1.2cm}|C{0.6cm}|}
\hline
 & Layer Type & Context factor 1 & Context factor 2 & Skip conn. from & Size & Inner size \\ \hline
1     & TDNN-ReLU    & t-2:t+2          &                  &                      & 512       &           \\ \hline
2     & FTDNN-ReLU   & t-2, t           & t+2, t           &                      & 512       & 256       \\ \hline
3     & FTDNN-ReLU   & t                & t                &                      & 512       & 256       \\ \hline
4     & FTDNN-ReLU   & t-3, t           & t-3, t           &                      & 512       & 256       \\ \hline
5     & FTDNN-ReLU   & t                & t                & 3                    & 512       & 256       \\ \hline
6     & FTDNN-ReLU   & t-3, t           & t+3, t           &                      & 512       & 256       \\ \hline
7     & FTDNN-ReLU   & t                & t                & 5                    & 512       & 256       \\ \hline
8     & FTDNN-ReLU   & t-3, t           & t+3, t           & 4                    & 512       & 256       \\ \hline
9     & FTDNN-ReLU   & t                & t                &                      & 512       & 256       \\ \hline
10    & TDNN-ReLU    & t                &                  &                      & 1536      &           \\ \hline
11    & Pooling &                  &                  &                      & 2*1000    &           \\ \hline
12    & Dense-ReLU    &                  &                  &                      & 512       &           \\ \hline
13    & Dense-ReLU    &                  &                  &                      & 512       &           \\ \hline
14    & Dense-Softmax &                  &                  &                      & N spkr &           \\ \hline
\end{tabular}
}
\end{table}
\subsubsection{C-vector based systems}
\label{cfix}

C-vector embedding architecture is based on residual blocks built using TDNN architecture, MFM (Max-Feature-Map) activations \cite{WHST2018} and A-Softmax (Angular Softmax) activation \cite{LWYLRS2017}.

\begin{figure}[]
	\centering
	\includegraphics[width=\columnwidth, trim=0 1.2cm 0 1.5cm]{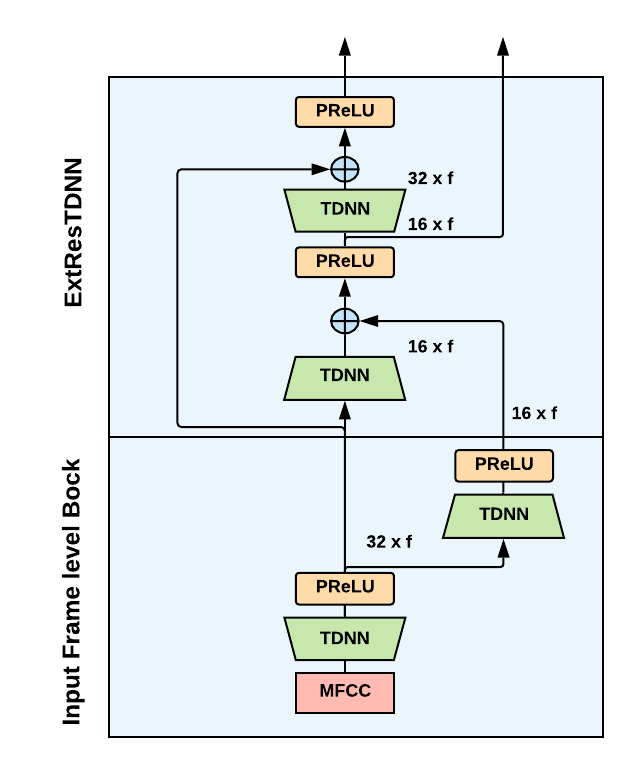}
	\caption{Residual block structure, used in C-vector based systems. Here \textbf{f} denotes the fixed parameter, which defines the size of layers used in Extended ResTDNN blocks.}\label{fig:Resblock}
\end{figure}

One of the proposed c-vector systems uses the original ResTDNN blocks from \cite{NSKKS2018}, while others utilize Extended TDNN blocks schematically described in Figure \ref{fig:Resblock}. The main differences between presented systems are the number of these Extended ResTDNN blocks and the value of the fixed parameter \textbf{(f)} that defines the size of layers used in these blocks. 

All systems use 40 dimensional MFCC with local CMN-normalization and global CMVN-normalization as input features and CSML as a backend.

\textbf{Cvec-ResTDNN-V1}: Original SpeakerResNet44 based extractor (c-vector) proposed in \cite{NSKKS2018} and trained on FixData-I. This architecture contains 20 basic ResTDNN blocks, described in \cite{NSKKS2018}, with one skip connection.

\textbf{Cvec-ExtResTDNN-V1}: This system contains 20 Extended ResTDNN blocks with fixed parameter f = 2. FixData-I was used for training.

\textbf{Cvec-ExtResTDNN-V2}: This system contains 26 Extended ResTDNN blocks with fixed parameter f = 2. FixData-II was used for training.

\textbf{Cvec-ExtResTDNN-V1-V2}: This system uses 20 Extended ResTDNN blocks with fixed parameter f = 2. FixData-I and FixData-II were used for training.

\textbf{Cvec-Wide-ResTDNN-V2}: This system is similar to previous one, it also contains 20 Extended ResTDNN blocks, but they are wider because of the fixed parameter \textbf{f} = 4. Only Database-II was used for training.

\textbf{Cvec-Wide-ExtResTDNN-V2}: This system contains 24 Extended ResTDNN blocks and fixed parameter \textbf{f} = 5. Only Database-II was used for training.


\subsection{Open training data conditions}
\label{open}
For  open  training  conditions along with individual systems from  Section \ref{common_fixed}, we used the following single systems trained on OpenData-IV:

\textbf{Xvec-TDNN-V4}: The configuration of this system is similar to Xvect-TDNN-V4 used for fixed conditions. In contrast, it uses 23 dimensional MFCC with local CMN and global CMVN-normalization.

\textbf{Xvec-TDNN-V4-WPE}: The embedding extractor in this system is the same as in Xvec-TDNN-V4, the only difference is that the test input speech signals were dereverberated by WPE algorithm before MFCC extraction.

\textbf{Cvec-ResTDNN-V4}: Original c-vector based system SpeakerResNet44 from \cite{NSKKS2018}. This system contains 20 basic ResTDNN blocks, described in \cite{NSKKS2018}, with one skip connection. 

\textbf{Cvec-ResTDNN-V1-WPE}: The embedding extractor in this system is the same as in Cvec-ResTDNN-V1, the test input speech signals were dereverberated by WPE algorithm before MFCC extraction.





\section{Submitted systems} \label{sec:performance}


For the fixed conditions all single systems described in sections \ref{xfixed} and \ref{cfix} were used. For the open conditions we used all fixed condition subsystems together with the subsystems described in \ref{open}. We used different score normalization and fusion strategies mentioned in \ref{fusion_calibration}:

\textbf{Fixed1 / Open1}:  the final score was estimated as the mean LLR score of two fused subsystems: 1) devset-II was used for the subsystems scores normalization, fusion was implemented with equal weights, devset-I was used for the system calibration; 2) devset-I was used for the subsystems scores normalization, fusion was implemented with equal weights, devset-II was used for the system calibration.

\textbf{Fixed2 / Open2}: devset-II was used for the subsystems scores normalization, fusion was implemented with equal weights, devset-I was used for the system calibration.

\textbf{Fixed3 / Open3}: devset-II was used for the subsystems scores normalization, fusion weights were obtained  empirically  (based  on  the  individual subsystems quality estimation), devset-II was used for the system calibration.

\
\section{Results and discussion}
Experiment results for our single and fusion systems on the development and evaluation sets are presented in  Tables~\ref{tab:table_results_single} and \ref{tab:table_results_fusion} respectively, in terms of EER (Equal Error Rate), minDCF (minimum Detection Cost Function), actDCF (actual Detection Cost Function) and Cllr (Log-Likelihood Ratio Cost) metrics using official scoring software \cite{nandwana2019voices}.
\begin{table}[h]
\caption{Results of our single systems on VOiCES 2019 challenge development set / evaluation set without score normalization}
\label{tab:table_results_single}
\resizebox{\columnwidth}{!}{
\begin{tabular}{l|c|c}
\hline
\textbf{System, fixed conditions}                            & \textbf{EER, \%}     & \textbf{minDCF} \\ \hline
Xvec-TDNN-V1                      & 3.01 / 8.55  & 0.276 / 0.552      \\
Xvec-TDNN-V2                       & 2.95 / 6.21 & 0.280 / 0.454      \\
Xvec-TDNN-V3             & 2.48 / 7.24  & 0.240 / 0.498      \\
Xvec-Ext-TDNN-V1                    & 2.52 / 6.00  & 0.242 / 0.395     \\
Xvec-Ext-TDNN-V3                     & 2.41 / 5.20  & 0.202 / 0.378      \\
Xvec-TDNN-LSTM-V1                    & 2.21 / 6.09  & 0.208 / 0.407      \\
Xvec-TDNN-LSTM-V3                    & 2.33 / 5.04  & 0.208 / 0.362      \\
Xvect-FTDNN-V2                         & 2.33 / 5.89  & 0.227 / 0.414      \\
Xvect-Ext-TDNN-LSTM-V3               & 2.56 / 5.16  & \textbf{0.194 / 0.349}      \\

Cvec-Wide-ResTDNN-V2     & 3.59 / 6.28  & 0.360 / \textbf{0.421}      \\
Cvec-Wide-ExtResTDNN-V2 & 3.44 / 6.64  & 0.327 / 0.456      \\ 
Cvec-ExtResTDNN-V1              & 3.35 / 6.63  & \textbf{0.269} / 0.459      \\
Cvec-ResTDNN-V1             & 4.28 / 7.03  & 0.372 / 0.506      \\
Cvec-ExtResTDNN-V1-V2       & 3.51 / 6.31  & 0.297 / 0.442      \\
Cvec-ExtResTDNN-V2               & 3.56 / 6.74  & 0.285 / 0.467      \\
\hline
\textbf{System, open conditions} & &\\
\hline
Xvec-TDNN-V4                     & 6.34 / 11.26 & 0.505 / 0.656      \\
Xvec-TDNN-V4-WPE                & 5.57 / 10.08 & \textbf{0.426 / 0.597}      \\
Cvec-ResTDNN-V4                   & 6.86 / 13.24 & 0.597 / 0.723      \\
Cvec-ResTDNN-V4-WPE              & 6.07 / 11.89 & \textbf{0.527 / 0.667}      \\
\end{tabular}}
\end{table}

\begin{table}[!h]
\centering
\caption{Results of our fused systems on VOiCES 2019 challenge development set / evaluation set with score normalization}
\label{tab:table_results_fusion}
\resizebox{\columnwidth}{!}{
\begin{tabular}{ccccc} 
\hline
\textbf{System} & \textbf{EER, \%} & \textbf{minDCF} & \textbf{actDCF} & \textbf{Cllr}
\\ 
\hline
Fixed-1 & 1.84 / 4.44 & 0.181 / \textbf{0.320} & 0.184 / \textbf{0.336}  & 0.066 / 0.290
\\
\hline
Fixed-2 & 1.84 / 4.49 & 0.181 / 0.324 & 0.184 / 0.342 & 0.066 / 0.264
\\
\hline
Fixed-3 & 1.87 / 4.51 & 0.187 / 0.323 & 0.190 / 0.342 & 0.068 / 0.261
\\
\hline
Open-1 & 1.69 / 4.44 & 0.177 / 0.320 & 0.182 / 0.334 & 0.065 / 0.203
\\
\hline
Open-2 & 1.62 / 4.53 & 0.172 / \textbf{0.315} & 0.200 / \textbf{0.320} & 0.175 / 0.216
\\
\hline
Open-3 & 1.82 / 4.49 & 0.181 / 0.320 & 0.184 / 0.342 & 0.067 / 0.260
\\
\hline
\end{tabular}}
\end{table}


It should be noted that deeper x-vector extractors with additional LSTM frame level layers perform better than original x-vector system.
Our best single system Xvect-Ext-TDNN-LSTM-V3 achieves top performance on both the development ($minDCF = 0.194$) and evaluation sets ($minDCF = 0.349$). 
According to the obtained results x-vector based systems are superior to the c-vector ones.
Additional attention should be paid to our results in open conditions: systems trained with the use of OpenData-IV demonstrate lower quality compared to those trained with FixData-[I,II,III]. Despite the increasing amount of training data, the downsampling from 16000 Hz to 8000 Hz leads to significant quality degradation of the considered systems.
We found out that for the SR task it is preferable to use 16000 Hz sampling rate and 40 dimensional MFCC features.

According to our observations the application of a more natural reverberation technique (like in FixData-[II, III]) for data augmentation makes the system more robust to unforeseen conditions. 
In some cases ASR based SAD (V3 systems) helps to achieve better quality than conventional energy-based SAD (V1, V2 systems).

WPE data preprocessing for data dereverberation improves systems for open conditions. It is important to note that preprocessing was applied only to the evaluation data but not during the training process.
It must be also pointed out that the main components of our SR systems are discriminatively trained models. With the use of the described systems, we were able to achieve the best quality values according to the Voices 2019 Challenge results.
Our studies also allow us to conclude that score s-normalization brings additional SR performance gain.

\section{Conclusions}

This paper demonstrates the efficiency of DNN-based speaker embedding extractors for speaker verification in single channel distant/far-field audio under noisy conditions. Deep extractors with additional LSTM frame-level layers before StatPooling layer allow improving SR systems quality. More realistic data augmentation procedure and the application of a powerful ASR-based SAD (FixData-III) lead to additional system performance improvements. Note that WPE dereverberation technique can be successfully implemented as an audio preprocessing step for the SR task.
The fusion of x-vector and c-vector based subsystems with CSML scoring model and scores s-normalization demonstrated the best performance on the Voices challenge data.

\bibliographystyle{IEEEtran}

\begin{thebibliography}{10}
\providecommand{\url}[1]{#1}
\csname url@samestyle\endcsname
\providecommand{\newblock}{\relax}
\providecommand{\bibinfo}[2]{#2}
\providecommand{\BIBentrySTDinterwordspacing}{\spaceskip=0pt\relax}
\providecommand{\BIBentryALTinterwordstretchfactor}{4}
\providecommand{\BIBentryALTinterwordspacing}{\spaceskip=\fontdimen2\font plus
\BIBentryALTinterwordstretchfactor\fontdimen3\font minus
  \fontdimen4\font\relax}
\providecommand{\BIBforeignlanguage}[2]{{%
\expandafter\ifx\csname l@#1\endcsname\relax
\typeout{** WARNING: IEEEtran.bst: No hyphenation pattern has been}%
\typeout{** loaded for the language `#1'. Using the pattern for}%
\typeout{** the default language instead.}%
\else
\language=\csname l@#1\endcsname
\fi
#2}}
\providecommand{\BIBdecl}{\relax}
\BIBdecl

\bibitem{sre_2016}
S.~O. Sadjadi, T.~Kheyrkhah, A.~Tong, C.~S. Greenberg, D.~A. Reynolds,
  E.~Singer, L.~P. Mason, and J.~Hernandez-Cordero, ``The 2016 nist speaker
  recognition evaluation.'' in \emph{Interspeech}, 2017, pp. 1353--1357.

\bibitem{NIST2018SREP}
\uppercase{NIST}, ``\uppercase{NIST} 2018 speaker recognition evaluation
  plan,'' https://www.nist.gov/file/453891, 2018, [Online; accessed
  03-October-2018].

\bibitem{SITW_ep}
M.~McLaren, L.~Ferrer, D.~Castan, and A.~Lawson, ``The 2016 speakers in the
  wild speaker recognition evaluation.'' in \emph{INTERSPEECH}, 2016, pp.
  823--827.

\bibitem{NSKKS2018}
S.~Novoselov, A.~Shulipa, I.~Kremnev, A.~Kozlov, and V.~Shchemelinin, ``On deep
  speaker embeddings for text-independent speaker recognition,'' in
  \emph{Odyssey 2018 The~Speaker and Language Recognition Workshop, June 26-29,
  Les Sables d'Olonne, France, Proceedings}, 2018, pp. 378--385.

\bibitem{huang2018angular}
Z.~Huang, S.~Wang, and K.~Yu, ``Angular softmax for short-duration
  text-independent speaker verification,'' \emph{Proc. Interspeech, Hyderabad},
  2018.

\bibitem{cai2018exploring}
W.~Cai, J.~Chen, and M.~Li, ``Exploring the encoding layer and loss function in
  end-to-end speaker and language recognition system,'' \emph{arXiv preprint
  arXiv:1804.05160}, 2018.

\bibitem{NSSKK2018}
S.~Novoselov, V.~Shchemelinin, A.~Shulipa, A.~Kozlov, and I.~Kremnev, ``Triplet
  loss based cosine similarity metric learning for text-independent speaker
  recognition,'' in \emph{INTERSPEECH 2018 -- 19\textsuperscript{th} Annual
  Conference of the International Speech Communication Association, August 2-6,
  Hyderabad, India, Proceedings}, 2018, pp. 2242--2246.

\bibitem{richey2018voices}
C.~Richey, M.~A. Barrios, Z.~Armstrong, C.~Bartels, H.~Franco, M.~Graciarena,
  A.~Lawson, M.~K. Nandwana, A.~Stauffer, J.~van Hout \emph{et~al.}, ``Voices
  obscured in complex environmental settings (voices) corpus,'' \emph{arXiv
  preprint arXiv:1804.05053}, 2018.

\bibitem{nandwana2019voices}
M.~K. Nandwana, J.~Van~Hout, M.~McLaren, C.~Richey, A.~Lawson, and M.~A.
  Barrios, ``The voices from a distance challenge 2019 evaluation plan,''
  \emph{arXiv preprint arXiv:1902.10828}, 2019.

\bibitem{MedennikovVOiCES2019}
I.~Medennikov, Y.~Khokhlov, A.~Romanenko, I.~Sorokin, A.~Mitrofanov, V.~Bataev,
  A.~Andrusenko, T.~Prisyach, M.~Korenevskaya, O.~Petrov, and A.~Zatvornitskiy,
  ``{The STC ASR System for the VOiCES from a Distance Challenge 2019},'' in
  \emph{INTERSPEECH (submitted)}, 2019.

\bibitem{Drude2018NaraWPE}
L.~Drude, J.~Heymann, C.~Boeddeker, and R.~Haeb-Umbach, ``{NARA-WPE: A Python
  package for weighted prediction error dereverberation in Numpy and Tensorflow
  for online and offline processing},'' in \emph{13. ITG Fachtagung
  Sprachkommunikation (ITG 2018)}, Oct 2018.

\bibitem{NYKMJ2010}
T.~Nakatani, T.~Yoshioka, K.~Kinoshita, M.~Miyoshi, and B.-H. Juang, ``Speech
  dereverberation based on variance-normalized delayed linear prediction,''
  \emph{IEEE Transactions on Audio, Speech, and Language Processing}, vol.~18,
  no.~7, pp. 1717--1731, 2010.

\bibitem{SGRSPK2018}
D.~Snyder, D.~Garcia-Romero, G.~Sell, D.~Povey, and S.~Khudanpur, ``X-vectors:
  Robust \uppercase{DNN} embeddings for speaker recognition,'' in
  \emph{ICASSP~2018 -- IEEE International Conference on Acoustics, Speech and
  Signal Processing, April 15-20, Calgary, Canada, Proceedings}, 2018, pp.
  5329--5333.

\bibitem{povey2011kaldi}
D.~Povey, A.~Ghoshal, G.~Boulianne, L.~Burget, O.~Glembek, N.~Goel,
  M.~Hannemann, P.~Motlicek, Y.~Qian, P.~Schwarz \emph{et~al.}, ``The kaldi
  speech recognition toolkit,'' in \emph{IEEE 2011 workshop on automatic speech
  recognition and understanding}, no. EPFL-CONF-192584.\hskip 1em plus 0.5em
  minus 0.4em\relax IEEE Signal Processing Society, 2011.

\bibitem{SGRPK2017}
D.~Snyder, D.~Garcia-Romero, D.~Povey, and S.~Khudanpur, ``Deep neural network
  embeddings for text-independent speaker verification,'' in \emph{INTERSPEECH
  2017 -- 18\textsuperscript{th} Annual Conference of the International Speech
  Communication Association, August 20-24, Stockholm, Sweden, Proceedings},
  2017, pp. 999--1003.

\bibitem{paszke2017automatic}
A.~Paszke, S.~Gross, S.~Chintala, G.~Chanan, E.~Yang, Z.~DeVito, Z.~Lin,
  A.~Desmaison, L.~Antiga, and A.~Lerer, ``{Automatic differentiation in
  PyTorch},'' in \emph{NIPS-W}, 2017.

\bibitem{NB2010}
H.~Nguyen and L.~Bai, ``Cosine similarity metric learning for face
  verification,'' in \emph{ACCV~2010 -- 10\textsuperscript{th} Asian Conference
  on Computer Vision, November 8-12, Queenstown, New Zealand, Proceedings},
  2010, pp. 709--720.

\bibitem{CVDFKCL2017}
D.~Colibro, C.~Vair, E.~Dalmasso, K.~Farell, G.~Karvitsky, S.~Cumani, and
  P.~Laface, ``Nuance--\uppercase{P}olitecnico di \uppercase{T}orino's 2016
  \uppercase{NIST} speaker recognition evaluation system,'' in
  \emph{INTERSPEECH 2017 -- 18\textsuperscript{th} Annual Conference of the
  International Speech Communication Association, August 20-24, Stockholm,
  Sweden, Proceedings}.\hskip 1em plus 0.5em minus 0.4em\relax International
  Speech Communication Association, 2017, pp. 1338--1342.

\bibitem{BV2013}
N.~Br{\"u}mmer and E.~de~Villiers, ``The \uppercase{BOSARIS} toolkit: Theory,
  algorithms and code for surviving the new \uppercase{DCF},'' in \emph{arXiv
  preprint arXiv:1304.2865}, 2013.

\bibitem{mclaren2018train}
M.~McLaren, D.~Castan, M.~K. Nandwana, L.~Ferrer, and E.~Yilmaz, ``How to train
  your speaker embeddings extractor,'' 2018.

\bibitem{AB1979}
J.~B. Allen and D.~A. Berkley, ``Image method for efficiently simulating
  small-room acoustics,'' \emph{The Journal of the Acoustical Society of
  America}, vol.~65, no.~4, pp. 943--950, 1979.

\bibitem{WHST2018}
X.~Wu, R.~He, Z.~Sun, and T.~Tan, ``A light \uppercase{CNN} for deep face
  representation with noisy labels,'' \emph{IEEE Transactions on Information
  Forensics and Security}, vol.~13, pp. 2884--2896, 2018.

\bibitem{LWYLRS2017}
W.~Liu, Y.~Wen, Z.~Yu, M.~Li, B.~Raj, and L.~Song, ``Sphereface: Deep
  hypersphere embedding for face recognition,'' in \emph{2017 IEEE Conference
  on Computer Vision and Pattern Recognition (CVPR~2017)}, vol.~1.\hskip 1em
  plus 0.5em minus 0.4em\relax IEEE, 2017, pp. 6738--6746.

\end{thebibliography}

\end{document}